\documentclass{llncs}

\usepackage[hyphens]{url}
\usepackage{times}
\usepackage{multicol}
\usepackage{amsmath}
\usepackage{amsfonts}
\usepackage{xfrac}
\usepackage{graphicx}
\usepackage{float}
\usepackage{rotating}
\usepackage{graphicx}
\usepackage{tabularx}
\usepackage{tabulary}
\usepackage{booktabs}
\usepackage{chngcntr}
\usepackage{xspace}
\counterwithout{table}{section}
\counterwithout{figure}{section}
\counterwithin{paragraph}{subsection}
\usepackage{multirow}
\usepackage{mdwlist}
\usepackage{paralist}
\usepackage{todonotes}
\usepackage{subcaption}
\usepackage{color}
\usepackage{listings}

\definecolor{lightgray}{rgb}{.6,.6,.6}
\definecolor{darkgray}{rgb}{.4,.4,.4}

\lstdefinelanguage{JSON}{
  keywords={break, case, catch, continue, debugger, default, delete, do, else, false, finally, for, function, if, in, instanceof, new, null, return, switch, this, throw, true, try, typeof, var, void, while, with},
  morecomment=[l]{//},
  morecomment=[s]{/*}{*/},
  morestring=[b]',
  morestring=[b]",
  ndkeywords={class, export, boolean, throw, implements, import, this},
  keywordstyle=\color{black}\bfseries,
  ndkeywordstyle=\color{darkgray}\bfseries,
  identifierstyle=\color{black},
  commentstyle=\color{lightgray}\ttfamily,
  stringstyle=\color{black}\ttfamily,
  sensitive=true
}

\lstdefinelanguage{JavaScript}{
  keywords={break, case, catch, continue, debugger, default, delete, do, else, false, finally, for, function, if, in, instanceof, new, null, return, switch, this, throw, true, try, typeof, var, void, while, with},
  morecomment=[l]{//},
  morecomment=[s]{/*}{*/},
  morestring=[b]',
  morestring=[b]",
  ndkeywords={class, export, boolean, throw, implements, import, this},
  keywordstyle=\color{black}\bfseries,
  ndkeywordstyle=\color{darkgray}\bfseries,
  identifierstyle=\color{black},
  commentstyle=\color{lightgray}\ttfamily,
  stringstyle=\color{red}\ttfamily,
  sensitive=true
}

\newcommand{\thesystem}{\textsc{OriginTracer}\xspace}

\begin{document}

\date{}

\title{Identifying Extension-based Ad Injection via Fine-grained Web Content Provenance}

\author{
  Sajjad Arshad, Amin Kharraz, and William Robertson\\
  Northeastern University, Boston, USA\\
  \texttt{\{arshad,mkharraz,wkr\}@ccs.neu.edu}
}

\institute{}

\maketitle

\begin{abstract}

Extensions provide useful additional functionality for web browsers, but are
also an increasingly popular vector for attacks. Due to the high degree of
privilege extensions can hold, extensions have been abused to inject
advertisements into web pages that divert revenue from content publishers and
potentially expose users to malware. Users are often unaware of such practices,
believing the modifications to the page originate from publishers. Additionally,
automated identification of unwanted third-party modifications is fundamentally
difficult, as users are the ultimate arbiters of whether content is undesired in
the absence of outright malice.

To resolve this dilemma, we present a fine-grained approach to tracking the
provenance of web content at the level of individual DOM elements. In
conjunction with visual indicators, provenance information can be used to
reliably determine the source of content modifications, distinguishing publisher
content from content that originates from third parties such as extensions. We
describe a prototype implementation of the approach called \thesystem for
Chromium, and evaluate its effectiveness, usability, and performance overhead
through a user study and automated experiments. The results demonstrate a
statistically significant improvement in the ability of users to identify
unwanted third-party content such as injected ads with modest performance
overhead.

\end{abstract}


\section{Introduction}
\label{sec:intro}

Browser extensions enhance browsers with additional useful capabilities that are
not necessarily maintained or supported by the browser vendor. Instead, this
code is typically written by third parties and can perform a wide range of
tasks, from simple changes in the appearance of web pages to sophisticated tasks
such as fine-grained filtering of content. To achieve these capabilities,
browser extensions possess more privilege than other third-party code that runs
in the browser. For instance, extensions can access cross-domain content, and
perform network requests that are not subject to the same origin policy. Because
these extensive capabilities allow a comparatively greater degree of control
over the browser, they provide a unique opportunity to attack users and their
data, the underlying system, and even the Internet at large. For this reason,
newer browser extension frameworks such as Chromium's have integrated least
privilege separation via isolated worlds and a fine-grained permissions system
to restrict the capabilities of third-party extensions%
~\cite{tr2008chromium-security-architecture}.

However, extension security frameworks are not a panacea. In practice, their
effectiveness is degraded by over-privilege and a lack of understanding of the
threats posed by highly-privileged extensions on the part of users%
~\cite{usenixwebapps2011app-permissions}. Indeed, despite the existence of
extension security frameworks, it has recently been shown that extension-based
advertisement injection has become a popular and lucrative technique for
dishonest parties to monetize user web browsing. These extensions simply inject
or replace ads in web pages when users visit a website, thus creating or
diverting an existing revenue stream to the third party. Users often are not
aware of these incidents and, even if this behavior is noticed, it can be
difficult to identify the responsible party.

While ad injection cannot necessarily be categorized as an outright malicious
activity on its own, it is highly likely that many users in fact \emph{do not
want or expect} browser extensions to inject advertisements or other content
into Web pages. Moreover, it can have a significant impact on the security and
privacy of both users as well as website publishers. For example, recent studies
have shown that ad-injecting extensions not only serve ads from ad networks
other than the ones with which the website publishers intended, but they also
attempt to trick users into installing malware by inserting rogue elements into
the web page~\cite{sp2015adinjection,www2015adinjection}.

To address this problem, several automatic approaches have been proposed to
detect malicious behaviors (e.g., ad injection) in browser extensions
\cite{www2015adinjection,usenixsec2014hulk,usenixsec2015webeval}. In addition,
centralized distribution points such as Chrome Web Store
and Mozilla Add-ons are using semi-automated techniques
for review of extension behavior to detect misbehaving extensions. However,
there is no guarantee that analyzing the extensions for a limited period of time
leads to revealing the ad injection behaviors. Finally, a client-side detection
methodology has been proposed in \cite{sp2015adinjection} that reports any
deviation from a legitimate DOM structure as potential ad injections. However,
this approach requires a priori knowledge of a legitimate DOM structure as well
as cooperation from content publishers.

Although ad injection can therefore potentially pose significant risks, this
issue is not as clear-cut as it might first seem. Some users might legitimately
want the third-party content injected by the extensions they install, even
including injected advertisements. This creates a fundamental dilemma for
automated techniques that aim to identify clearly malicious or unwanted content
injection, since such techniques cannot intuit user intent and desires in a
fully automatic way.

To resolve this dilemma, we present \thesystem, an in-browser approach to
highlight extension-based content modification of web pages. \thesystem monitors
the execution of browser extensions to detect content modifications such as the
injection of advertisements. Content modifications are visually highlighted in
the context of the web page in order to
\begin{inparaenum}[\itshape i)\upshape]
     \item notify users of the presence of modified content, and
     \item inform users of the \emph{source} of the modifications.
\end{inparaenum}
With this information, users can then make an informed decision as to whether
they actually want these content modifications from specific extensions, or
whether they would rather uninstall the extensions that violate their
expectations.

\thesystem assists users in detecting content injection by distinguishing
injected or modified DOM elements from genuine page elements. This is performed
by annotating web page DOM elements with a \emph{provenance label set} that
indicates the principal(s) responsible for adding or modifying that element,
both while the page is loading from the publisher as well as during normal
script and extension execution. These annotations serve as trustworthy,
fine-grained provenance indicators for web page content. \thesystem can be
easily integrated into any browser in order to inform users of extension-based
content modification. Since, \thesystem identifies all types of content
injections, it is able to highlight all injected advertisements regardless of
their types (e.g., flash ads, banner ads, and text ads).

We implemented a prototype of \thesystem as a set of modifications to the
Chromium browser, and evaluated its effectiveness by conducting a user study.
The user study reveals that \thesystem produced a significantly greater
awareness of third-party content modification, and did not detract from the
users' browsing experience. Our results also suggests that \thesystem can be
used as a complementary system to ad blocking systems such as Adblock Plus%
~\cite{adblock-plus} and Ghostery~\cite{ghostery}.

To summarize, the main contributions of this paper are:

\begin{itemize*}

\item We introduce a novel in-browser approach to provenance tracking for web
content at the granularity of DOM elements, and present a semantics for
provenance propagation due to script and extension execution. The approach
leverages a high-fidelity in-browser vantage point that allows it to construct a
precise provenance label set for each DOM element introduced into a web page.

\item We implement a prototype called \thesystem that uses content provenance to
identify and highlight third-party content injection -- e.g., unwanted
advertisements -- by extensions to notify users of their presence and the
originating principal.

\item We evaluate the effectiveness, usability, and performance of our
prototype, and show that it is able to significantly assist users in identifying
ad injection by extensions in the wild without degrading browser performance or
the user experience.

\end{itemize*}

\section{Background \& Motivation}
\label{sec:background-motivation}

In the following, we introduce background information on browser extensions,
present an overview of advertisement injection as a canonical example of
questionable content modification, and motivate our approach in this context.

\subsection{Browser Extensions}
\label{sec:background-motivation:extensions}

Browser extensions are programs that extend the functionality of a web browser.
Today, extensions are typically implemented using a combination of HTML, CSS,
and JavaScript written against a browser-specific extension API. These APIs
expose the ability to modify the browser user interface in controlled ways,
manipulate HTTP headers, and modify web page content through the document object
model (DOM) API. An extension ecosystem is provided by almost all major browser
vendors; for instance, Google and Mozilla both host centralized repositories of
extensions that users can download at the Chrome Web Store and Mozilla Add-ons
sites, respectively.

\subsection{Advertisement Injection}
\label{sec:background-motivation:ad-injection}

\begin{figure}[t]
    {\sffamily
    \centering
    \includegraphics[width=.6\linewidth]{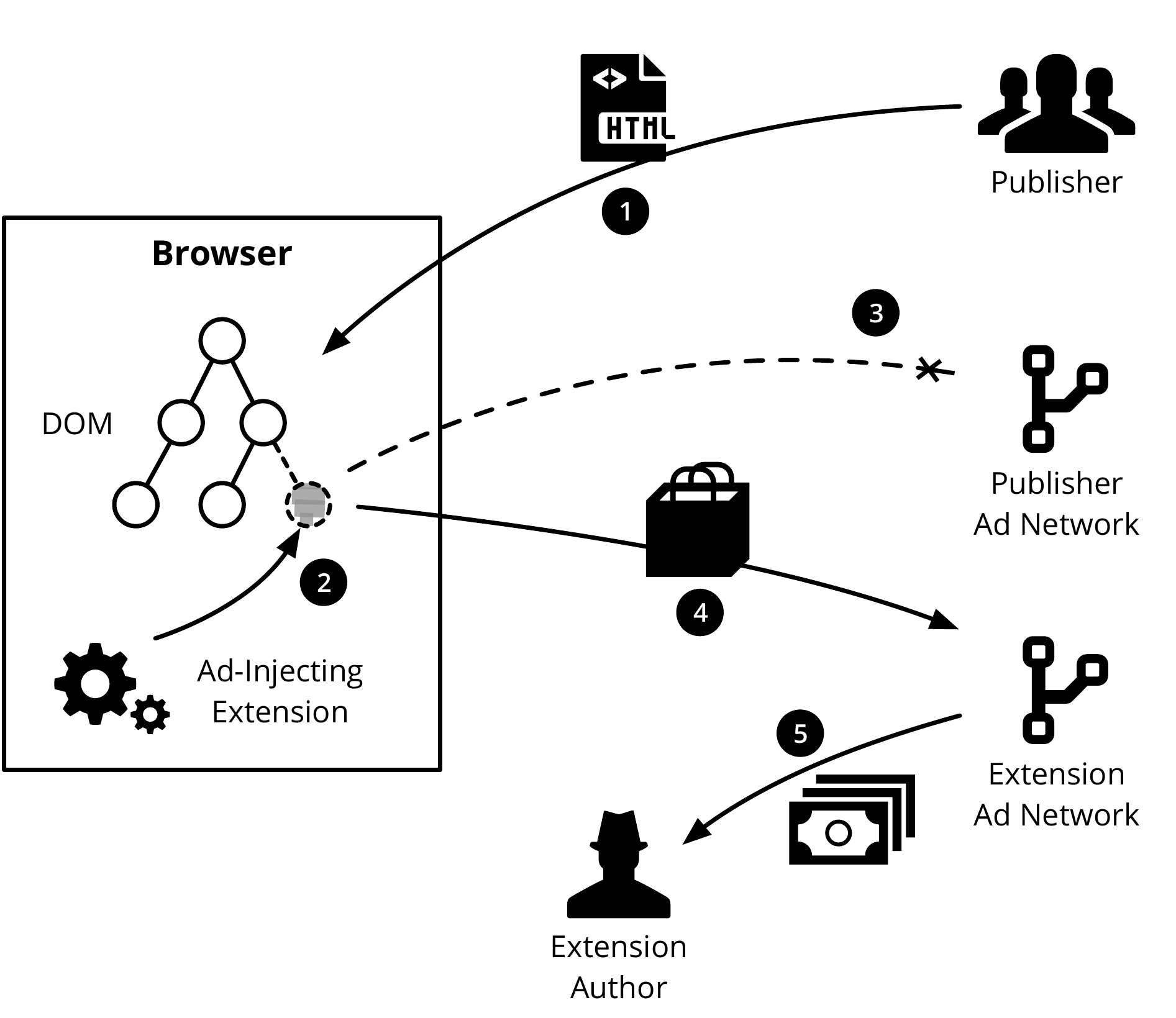}
    \caption{Overview of advertisement injection.
        \textbf{(1)}~The user accesses the publisher's site.
        \textbf{(2)}~An ad-injecting browser extension adds DOM elements to display ads to the user, and optionally removes existing ads.
        \textbf{(3)}~Ad revenue is diverted from the publisher.
        \textbf{(4)}~Ad impressions, clicks, and conversions are instead directed to the extension's ad network.
        \textbf{(5)}~Ad revenue flows to the extension author.
    }
    \vspace{-2.1em}
    \label{fig:ad-injection}
    }
\end{figure}

As web advertising grew in popularity, those in a position to modify web content
such as ISPs and browser extension authors realized that profit could be
realized by injecting or replacing ads in web pages. For instance, some ISPs
began to tamper with HTTP traffic in transit, injecting DOM elements into HTML
documents that added ISP's advertisements into pages visited by their
customers~\cite{isp-adinjection2,isp-adinjection1}. In a similar fashion,
browser extensions started modifying pages to inject DOM elements in order to
show ads to users without necessarily obtaining the user's prior consent. Ad
injection has evolved to become a common form of unrequested third-party content
injection on today's web~\cite{adinjection-profit}.

These practices have several effects on both publishers and users. On one hand,
ad injection diverts revenue from the publisher to the third party responsible
for the ad injection. If advertisements are the primary source of income for a
publisher, this can have a significant effect on their bottom line. If the
injected ads contain or reference undesired content (e.g., adult or political
topics), ad injection can also harm the reputation of the publisher from the
user's perspective. If the content injection is also malicious in nature, the
publisher's reputation can be further harmed in addition to exposing users to
security risks due to malware, phishing, and other threats. Prior work has shown
that users exposed to ad injection are more likely to be exposed to
``malvertising'' and traditional malware~%
\cite{sp2015adinjection,www2015adinjection}. Figure~\ref{fig:ad-injection}
gives an overview of ad injection's effect on the normal ad delivery process,
while Figure~\ref{fig:indicator} shows an instance of ad injection on
amazon.com.

\subsection{Motivation}
\label{sec:background-motivation:motivation}

\begin{table}[t]
    \centering
    \setlength{\tabcolsep}{22pt}
    {\sffamily\fontsize{9}{11}\selectfont
    \begin{tabularx}{\linewidth}{lrc}
    \toprule
    \textbf{Extension} & \textbf{No. of Users} & \textbf{Injected Element} \\
    \midrule
    Adblock Plus & 10,000,000+ & \texttt{\textless iframe\textgreater} \\
    Google Translate & 6,000,000+ & \texttt{\textless div\textgreater} \\
    Tampermonkey & 5,800,000+ & \texttt{\textless img\textgreater} \\
    Evernote Web Clipper & 4,300,000+ & \texttt{\textless iframe\textgreater} \\
    Google Dictionary & 3,000,000+ & \texttt{\textless div\textgreater} \\
    \bottomrule
    \end{tabularx}
    \caption{Five popular Chrome extensions that modify web pages as part of
    their benign functionality.}
    \vspace{-2.5em}
    \label{tab:popular-extensions}
    }
\end{table}

Recently, there have been efforts by browser vendors to remove ad-injecting
extensions from their repositories~\cite{google-adinjection}. Although
semi-automated central approaches have been successful in identifying
ad-injecting extensions, deceptive extensions can simply hide their ad injection
behaviors during the short period of analysis time. In addition, finding web
pages that trigger ad injection is a non-trivial task, and they can miss some
ad-injecting extensions. Moreover, there are extensions that are not provided
through the web stores, and users can get them from local marketplaces, which
may not examined the extensions properly. Hence, we believe that there is a need
for a protection tool to combat ad injection on the client side in addition to
centralized examination by browser vendors.

Furthermore, automatically determining whether third-party content modification
-- such as that due to ad injection -- should be allowed is not straightforward.
Benign extensions extensively modify web pages as part of their normal
functionality. To substantiate this, we examined five popular Chrome extensions
as of the time of writing; these are listed in Table~%
\ref{tab:popular-extensions}. Each of these extensions are available for all
major browsers, and all modify web pages (e.g., inject elements) to implement
their functionality. Therefore, automated approaches based on this criterion run
a high risk of false positives when attempting to identify malicious or
undesirable extensions.

Moreover, it is not enough to identify that advertisements, for instance, have
been injected by a third party. This is because some users \emph{might
legitimately desire} the content that is being added to web pages by the
extensions they install. To wit, it is primarily for this reason that a recent
purge of extensions from the Chrome Web Store did not encompass the entirety of
the extensions that were identified as suspicious in a previous study, as the
third-party content modification could not be clearly considered as
malicious~\cite{sp2015adinjection}. Instead, we claim that \emph{users
themselves} are best positioned to make the determination as to whether
third-party content modification is desired or not. An approach that proceeds
from this observation would provide sufficient, easily comprehensible
information to users in order to allow an informed choice as to whether content
is desirable or should be blocked. It should be noted that defending against
drive-by downloads and general malware is not the focus of this paper. Rather,
the goal is to highlight injected ads to increase likelihood that user will make
an informed choice to not click on them.

We envision that \thesystem could be used as a complementary approach to
existing techniques such as central approaches used by browser vendors. Also,
browser vendors can benefit from using our system in addition to end users to
detect the content modifications by extensions in a more precise and reliable
way. In the following sections, we present design and implementation of our
system.

\section{Web Content Provenance}
\label{sec:design}

In this section, we describe an in-browser approach for identifying third-party
content modifications in web browsers. The approach adds \emph{fine-grained
provenance tracking} to the browser, at the level of individual DOM elements.
Provenance information is used in two ways:
\begin{inparaenum}[\itshape i)\upshape]
    \item to distinguish between content that originates from the web page
    publisher and content injected by an unassociated third party, and
    \item to indicate \emph{which} third party (e.g., extension) is responsible
    for content modifications using provenance indicators.
\end{inparaenum}
By integrating the approach directly into the browser, we guarantee the
trustworthiness of both the provenance information and the visual indicators.
That is, as the browser is already part of the trusted computing base (TCB) in
the web security model, we leverage this as the appropriate layer to compute
precise, fine-grained provenance information. Similarly, the browser holds
sufficient information to ensure that provenance indicators cannot be tampered
with or occluded by malicious extensions. While we consider malicious or
exploited browser plug-ins such as Flash Player outside our threat model, we
note that modern browsers take great pains to isolate plug-ins in least
privilege protection domains. We report separately on the implementation of
the approach in Section~\ref{sec:impl}.

In the following, we present our approach to tracking and propagating content
provenance, and then discuss provenance indicators and remediation strategies.

\subsection{Content Provenance}
\label{sec:design:provenance}

\begin{figure}[t]
    {\sffamily
    \centering
    \includegraphics[width=.7\linewidth]{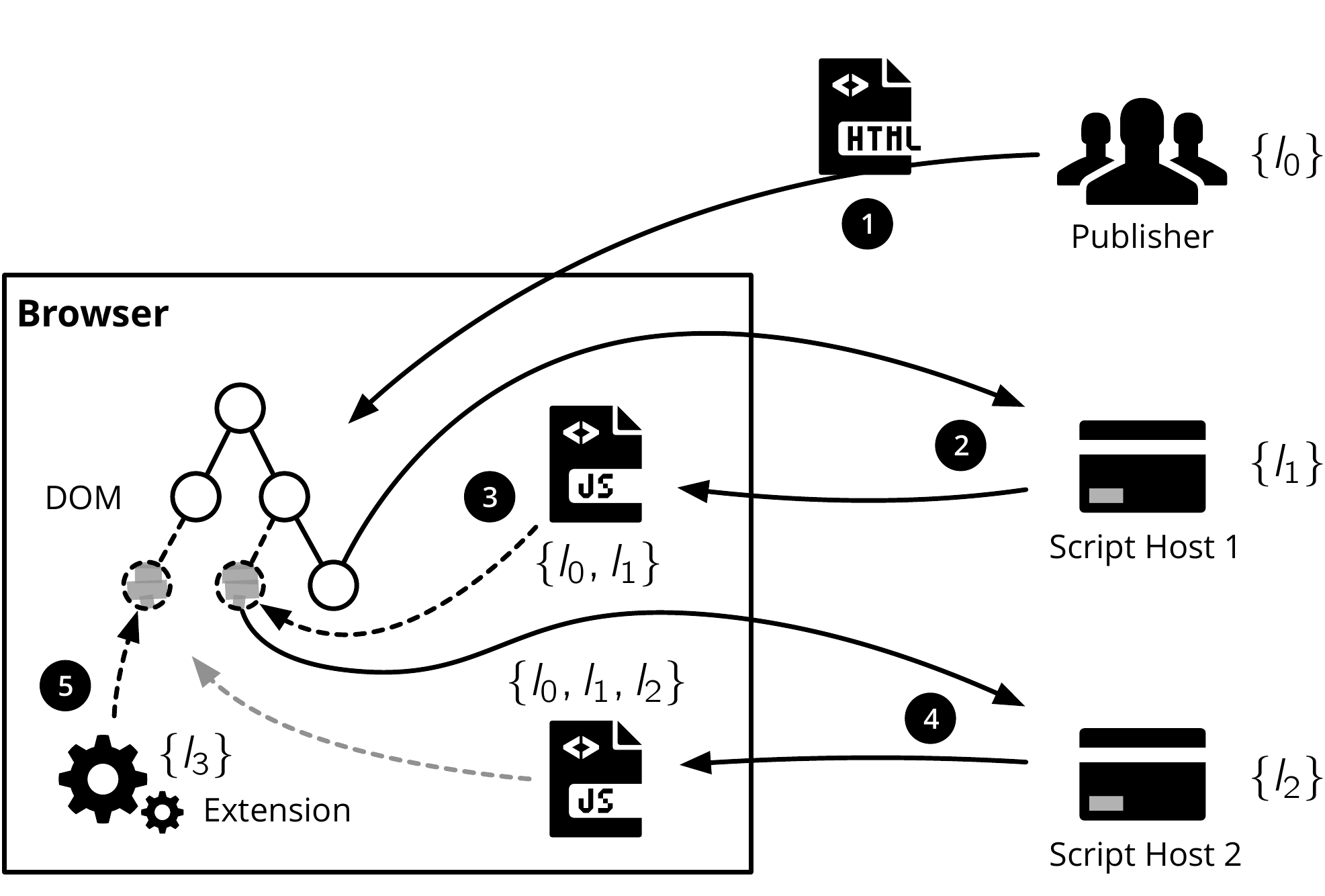}
    \caption{Element-granularity provenance tracking.
        \textbf{(1)}~Content loaded directly from the publisher is labeled
        with the publisher's origin, \(l_0\).
        \textbf{(2)}~An external script reference to origin \(l_1\) is performed.
        \textbf{(3)}~DOM modifications from \(l_1\)'s script are labeled with
        the label set \(\left\{l_0,l_1\right\}\).
        \textbf{(4)}~Further external script loads and subsequent DOM
        modifications induce updated label sets -- e.g.,
        \(\left\{l_0,l_1,l_2\right\}\).
        \textbf{(5)}~A DOM modification that originates from an extension
        produces provenance label sets \(\left\{l_0, l_1, l_2, l_3\right\}\) for
        the element.}
    \vspace{-1.5em}
    \label{fig:provenance}
    }
\end{figure}

Web pages are composed of HTML that references resources such as stylesheets,
scripts, images, plug-ins such as Flash objects, or even other web pages loaded
inside frames. The document object model (DOM) is a natural
structural representation of a web page that can be manipulated through a
standard API, and serves as a suitable basis for provenance tracking. In
particular, our system tracks the provenance of each element \(e\) contained in
a DOM. Provenance for a DOM element is recorded as a set of labels \(\ell \in
\mathcal{P}\left(L\right)\), where the set of all labels \(L\) corresponds to a
generalization of standard web origins to include extensions. That is, instead
of the classic origin 3-tuple of \(\left\langle \textsf{scheme}, \textsf{host},
\textsf{port} \right\rangle\), we record
\begin{align*}
    L &= \left\langle S, I, P, X \right\rangle \\
    S &= \left\{ \textsf{scheme} \right\} \cup \left\{ \text{``extension''} \right\} \\
    I &= \left\{ \textsf{host} \right\} \cup \left\{ \textsf{extension-identifier} \right\} \\
    P &= \left\{ \textsf{port} \right\} \cup \left\{ \textsf{null} \right\} \\
    X &= \left\{0, 1, 2, \ldots\right\}
\end{align*}
In other words, a label is a 4-tuple that consists of a normal network scheme or
\textsf{extension}, a network host or a unique extension identifier, a port or
the special \textsf{null} value, and an index used to impose a global total
order on labels as described below. While browsers use different extension
identifiers, including randomly-generated identifiers, the exact representation
used is unimportant so long as there is a one-to-one mapping between extensions
and identifiers and their use is locally consistent within the browser. An
overview of provenance tracking is depicted in Figure~\ref{fig:provenance}.

\textbf{Static Publisher Provenance.}
Content provenance tracking begins with a web page load. As the DOM is parsed by
the browser, each element is labeled with a singleton label set containing the
origin of the publisher, \(\left\{l_0\right\}\). Thus, static provenance
tracking is straightforward and equivalent to the standard use of origins as a
browser security context.

\textbf{Dynamic Publisher Provenance.}
Content provenance becomes more interesting in the presence of dynamic code
execution. As JavaScript can add, modify, and remove DOM elements in an
arbitrary fashion using the DOM API exposed by the browser, it is necessary to
track these modifications in terms of provenance labels.

New provenance labels are created from the publisher's label set
\(\left\{l_0\right\}\) as follows. Whenever an external script is referenced
from the initial DOM resulting from the page load, a new label \(l_i, i \in
\left\{1,2,\ldots\right\}\) is generated from the origin of the script. All
subsequent DOM modifications that occur as a result of an external script loaded
from the initial DOM are recorded as \(\left\{l_0,l_i\right\}\). Successive
external script loads follow the expected inductive label generation process --
i.e., three successive external script loads from unique origins will result in
a label set \(\left\{l_0,l_i,l_j,l_k\right\}\). Finally, label sets contain
unique elements such that consecutive external script loads from a previously
accessed origin are not reflected in the label for subsequent DOM modifications.
For instance, if the web page publisher loads a script from the publisher's
origin, then any resulting DOM modifications will have a provenance label set of
\(\left\{l_0\right\}\) instead of \(\left\{l_0,l_0\right\}\). Content provenance
is propagated for three generic classes of DOM operations: element insertion,
modification, and deletion.

Element insertions produce an updated DOM that contains the new element labeled
with the current label set, and potentially generates a new label set if the
injected element is a script. Element modifications produce a DOM where the
modified element's label set is merged with the current label set. Finally,
element deletions simply remove the element from the DOM.

\textbf{Extension Provenance.}
The third and final form of provenance tracking concerns content modifications
due to DOM manipulations by extensions. In this case, provenance propagation
follows the semantics for the above case of dynamic publisher provenance. Where
these two cases differ, however, is in the provenance label initialization.
While provenance label sets for content that originates, perhaps indirectly,
from the web page publisher contains the publisher's origin label \(l_0\),
content that originates from an extension is rooted in a label set initialized
with the \emph{extension's} label. In particular, content modifications that
originate from an extension \emph{are not labeled} by the publisher's origin. An
exception to this occurs when the extension, either directly or indirectly,
subsequently loads scripts from the publisher, or modifies an existing element
that originated from the publisher.

\subsection{Content Provenance Indicators}
\label{sec:design:indicators}

With the fine-grained content provenance scheme described above, identifying the
principal responsible for DOM modifications is straightforward. For each
element, all that is required is to inspect its label set \(\ell\) to check
whether it contains the label of any extension.

A related, but separate, question is how best to relay this information to the
user. In this design, several options are possible on a continuum from simply
highlighting injected content without specific provenance information to
reporting the full ordered provenance chain from the root to the most recent
origin. The first option makes no use of the provenance chain, while the other
end of the spectrum is likely to overwhelm most users with too much information,
degrading the practical usefulness of provenance tracking. We suspect that a
reasonable balance between these two extremes is a summarization of the full
chain, for instance by reporting only the label of the corresponding extension.

Finally, if a user decides that the third-party content modification is
unwanted, another design parameter is how to act upon this decision. Possible
actions include blocking specific element modifications, removing the offending
extension, or reporting its behavior to a central authority. We report on the
specific design choices we made with respect to provenance indicators in the
presentation of our implementation in Section~\ref{sec:impl}.

\section{OriginTracer}
\label{sec:impl}

In this section, we present \thesystem, our prototype implementation for
identifying and highlighting extension-based web page content modifications. We
implemented \thesystem as a set of modifications to the Chromium browser. In
particular, we modified both Blink and the extension engine to track the
provenance of content insertion, modification, and removal according to the
semantics presented in Section~\ref{sec:design}. These modifications also
implement provenance indicators for suspicious content that does not originate
from the publisher. In total, our changes consist of approximately 900 SLOC for
C++ and several lines of JavaScript\footnote{SLOC were measured using David
Wheeler's SLOCCount~\cite{sloccount}.}. In the following, we provide more detail
on the integration of \thesystem into Chromium.

\subsection{Tracking Publisher Provenance}

A core component of \thesystem is responsible for introducing and propagating
provenance label sets for DOM elements. In the following, we discuss the
implementation of provenance tracking for publisher content.

\textbf{Tracking Static Elements.}
As discussed in Section~\ref{sec:design}, provenance label sets for static DOM
elements that comprise the HTML document sent by the publisher as part of the
initial page load are equivalent to the publisher's web origin -- in our
notation, \(l_0\). Therefore, minimal modifications to the HTML parser were
necessary to introduce these element annotations, which is performed in an
incremental fashion as the page is parsed.

\textbf{Tracking Dynamic Elements.}
To track dynamic content modifications, this component of \thesystem must also
monitor JavaScript execution. When a \texttt{script} tag is encountered during
parsing of a page, Blink creates a new element and attaches it to the DOM. Then,
Blink obtains the JavaScript code (fetching it from network in the case of
remote script reference), submits the script to the V8 JavaScript engine for
execution, and pauses the parsing process until the script execution is
finished. During execution of the script, some new elements might be created
dynamically and inserted into the DOM. According to the provenance semantics,
these new elements inherit the label set of the script. In order to create new
elements in JavaScript, one can
\begin{inparaenum}[\itshape i)\upshape]
    \item use DOM APIs to create a new element and attach it to the web page's
    DOM, or
    \item write HTML tags directly into the page.
\end{inparaenum}

In the first method, to create a new element object, a canonical example is to
provide the tag name to the \texttt{createElement} function. Then, other
attributes of the newly created element are set -- e.g., after creating an
element object for an \texttt{a} tag, an address must be provided for its
\texttt{href} attribute. Finally, the new element should be attached to the DOM
tree as a child using \texttt{appendChild} or \texttt{insertBefore} functions.
In the second method, HTML is inserted directly into the web page using the
functions such as \texttt{write} and \texttt{writeln}, or by modifying the
\texttt{innerHTML} attribute. In cases where existing elements are modified
(e.g., changing an image's \texttt{src} attribute), the element inherits the
label set of the currently executing script as well. In order to have a complete
mediation of all DOM modifications to Web page, \texttt{Node} class in Blink
implementation was instrumented in order to assign provenance label sets for
newly created or modified elements using the label set applied to the currently
executing script.

\textbf{Handling Events and Timers.}
An additional consideration for this \thesystem component is modifications to
event handlers and timer registrations, as developers make heavy use of event
and timer callbacks in modern JavaScript. For instance, such callbacks are used
to handle user interface events such as clicking on elements, hovering over
elements, or to schedule code after a time interval has elapsed. In practice,
this requires the registration of callback handlers via
\texttt{addEventListener} API for events, and \texttt{setTimeout} and
\texttt{setInterval} for timers. To mediate callbacks related to the addition
and firing of events and timers, we slightly modified the \texttt{EventTarget}
and \texttt{DOMTimer} classes in Blink, respectively. Specifically, we record
the mapping between the running scripts and their registered callback functions,
and then recover the responsible script for DOM modification during callback
execution.

\subsection{Tracking Extension Provenance}
\label{sec:impl:extension}

Chromium's extension engine is responsible for loading extensions, checking
their permissions against those declared in the manifest file, injecting content
scripts, dispatching background scripts and content scripts to the V8 script
engine for execution, and providing a channel for communication between content
scripts and background page.

Chromium extensions can manipulate the web page's content by injecting
\emph{content scripts} into the web page or using the \texttt{webRequest}
API. Content scripts are JavaScript programs that can manipulate the web page
using the shared DOM, communicate with external servers via
\texttt{XMLHttpRequest}, invoke a limited set of \texttt{chrome.*} APIs, and
interact with their owning extension's background pages. By using
\texttt{webRequest}, extensions are also able to modify and block HTTP
requests and responses in order to change the web page's DOM.

In this work, we only track content modifications by content scripts and leave
identifying ad injection by \texttt{webRequest} for future engineering work.
Prior work, however, has mentioned that only 5\% of ad injection incidents
occurred via \texttt{webRequest}; instead, Chrome extensions mostly rely on
content scripts to inject advertisements~\cite{sp2015adinjection}. Moreover,
with modern websites becoming more complex, injecting stealthy advertisement
into the page using \texttt{webRequest} is not a trivial task.

\textbf{Tracking Content Script Injection and Execution.}
To track elements created or modified during the execution of content scripts,
extension engine was modified to hook events corresponding to script injection
and execution. Content scripts can be inserted into the web page using different
methods. If a content script should be injected into every matched web page, it
must be registered in the extension manifest file using the
\texttt{content\_scripts} field. By providing different options for this field,
one can control when and where the content scripts be injected. Another method
is programmatic injection, which is useful when content scripts should be
injected in response to specific events (e.g., a user clicks the extension's
browser action). With programmatic injection, content scripts can be injected
using the \texttt{tabs.executeScript} API if the \texttt{tabs} permission is set
in the manifest file. Either way, content scripts have a provenance label set
initialized with the extension's label upon injection.

\textbf{Handling Callback Functions.}
Chromium's extension engine provides a messaging API as a communication channel
between background pages and content scripts. Background pages and content
scripts can receive messages from each other by providing a callback function
for the \texttt{onMessage} or \texttt{onRequest} events, and can send messages
by invoking \texttt{sendMessage} or \texttt{sendRequest}. To track the
registration and execution of callback functions, the \texttt{send\_request} and
\texttt{event} modules were slightly modified in the extension engine.
Specifically, we added some code to map registered callbacks to their
corresponding content scripts in order to find the extension responsible for DOM
modification.

\subsection{Content Provenance Indicators}
\label{sec:impl:indicators}

\begin{figure}[!t]
    \centering
    {\sffamily
    \includegraphics[width=.6\linewidth]{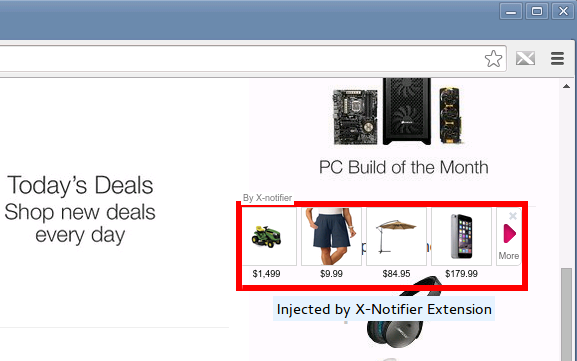}
    \caption{An example of indicator for an injected advertisement on amazon.com.}
    \vspace{-1.5em}
    \label{fig:indicator}
    }
\end{figure}

Given DOM provenance information, \thesystem must first
\begin{inparaenum}[\itshape i)\upshape]
    \item identify when suspicious content modifications -- e.g.,
    extension-based ad injection -- has occurred, and additionally
    \item communicate this information to the user in an easily comprehensible
    manner.
\end{inparaenum}
To implement the first requirement, our prototype monitors for content
modifications where a subtree of elements are annotated with label sets that
contains a particular extension's label. This check can be performed efficiently
by traversing the DOM and inspecting element label sets after a set of changes
have been performed on the DOM.

There are several possible options to communicate content provenance as
mentioned in Section~\ref{sec:design}. In our current prototype, provenance is
indicated using a configurable border color of the root element of the
suspicious DOM subtree. This border should be chosen to be visually distinct
from the existing color palette of the web page. Finally, a tooltip indicating
the root label is displayed when the user hovers their mouse over the DOM
subtree. An example is shown in Figure~\ref{fig:indicator}. To implement these
features, \thesystem modifies \texttt{style} and \texttt{title} attributes. In
addition, since \thesystem highlights elements in an online fashion, it must
delay the addition of highlighting until the element is attached to the page's
DOM and is displayed. Therefore, modifications were made to the
\texttt{ContainerNode} class that is responsible for attaching new elements to
the DOM.

While we did not exhaustively explore the design space of content provenance
indicators in this work (e.g., selective blocking of extension-based DOM
modifications), we report on the usability of the prototype implementation in
our evaluation.

\section{Evaluation}
\label{sec:eval}

In this section, we measure the effectiveness, usability, and performance of
content provenance indicators using the \thesystem prototype. In particular, the
questions we aim to answer with this evaluation are:

\begin{enumerate*}
    \item[\textbf{(Q1)}] \label{eval:q1}
    How susceptible are users to injected content such as third-party
    advertisements? (\S\ref{sec:eval:susceptibility})
    \item[\textbf{(Q2)}] \label{eval:q2}
    Do provenance indicators lead to a significant, measurable decrease in the
    likelihood of clicking on third-party content that originates from
    extensions? (\S\ref{sec:eval:effectiveness})
    \item[\textbf{(Q3)}] \label{eval:q3}
    Are users likely to use the system during their normal web browsing?
    (\S\ref{sec:eval:usability})
    \item[\textbf{(Q4)}] \label{eval:q4}
    Does integration of the provenance tracking system significantly degrade
    the users' browsing experience and performance of the browser on a
    representative sample of websites? (\S\ref{sec:eval:perf})
\end{enumerate*}

\textbf{Ethics Statement.} As part of the evaluation, we performed two
experiments involving users unaffiliated with the project as described below.
Due to the potential risk to user confidentiality and privacy, we formulated an
experimental protocol that was approved by our university's institutional review
board (IRB). This protocol included safeguards designed to prevent exposing
sensitive user data such as account names, passwords, personal addresses, and
financial information, as well as to protect the anonymity of the study
participants with respect to data storage and reporting. While users were not
initially told the purpose of some of the experiments, all users were debriefed
at the end of each trial as to the true purpose of the study.

\subsection{Effectiveness of the Approach}

Similar to prior work~\cite{chi2006phishing}, we performed a user study to
measure the effectiveness of content provenance in enabling users to more easily
identify unwanted third-party content. However, we performed the user study with
a significantly larger group of participants. The study population was composed
of 80 students that represent a range of technical sophistication. We conducted
an initial briefing prior to the experiments where we made it clear that we were
interested in honest answers.

\textbf{User Susceptibility to Ad Injection.}
\label{sec:eval:susceptibility}
The goal of the first phase of the experiment was to measure whether users were
able to detect third-party content that was not intended for inclusion by the
publishers of web pages presented to them. Users were divided into two equal
sized groups of 40. In each group, users were first presented with three
unmodified Chromium browsers, each of which had a separate ad-injecting
extension installed: \textsf{Auto Zoom}, \textsf{Alpha Finder}, and
\textsf{X-Notifier} for the first group, and \textsf{Candy Zapper},
\textsf{uTorrent}, and \textsf{Gethoneybadger} for the second group. These
extensions were chosen because they exhibit a range of ad injection behaviors,
from subtle injections that blend into the publisher's web page to very obvious
pop-ups that are visually distinct from the publisher's content.

\begin{figure}[t]
    \centering
    {\sffamily
    \begin{subfigure}[t]{0.48\linewidth}
        \includegraphics[width=\textwidth]{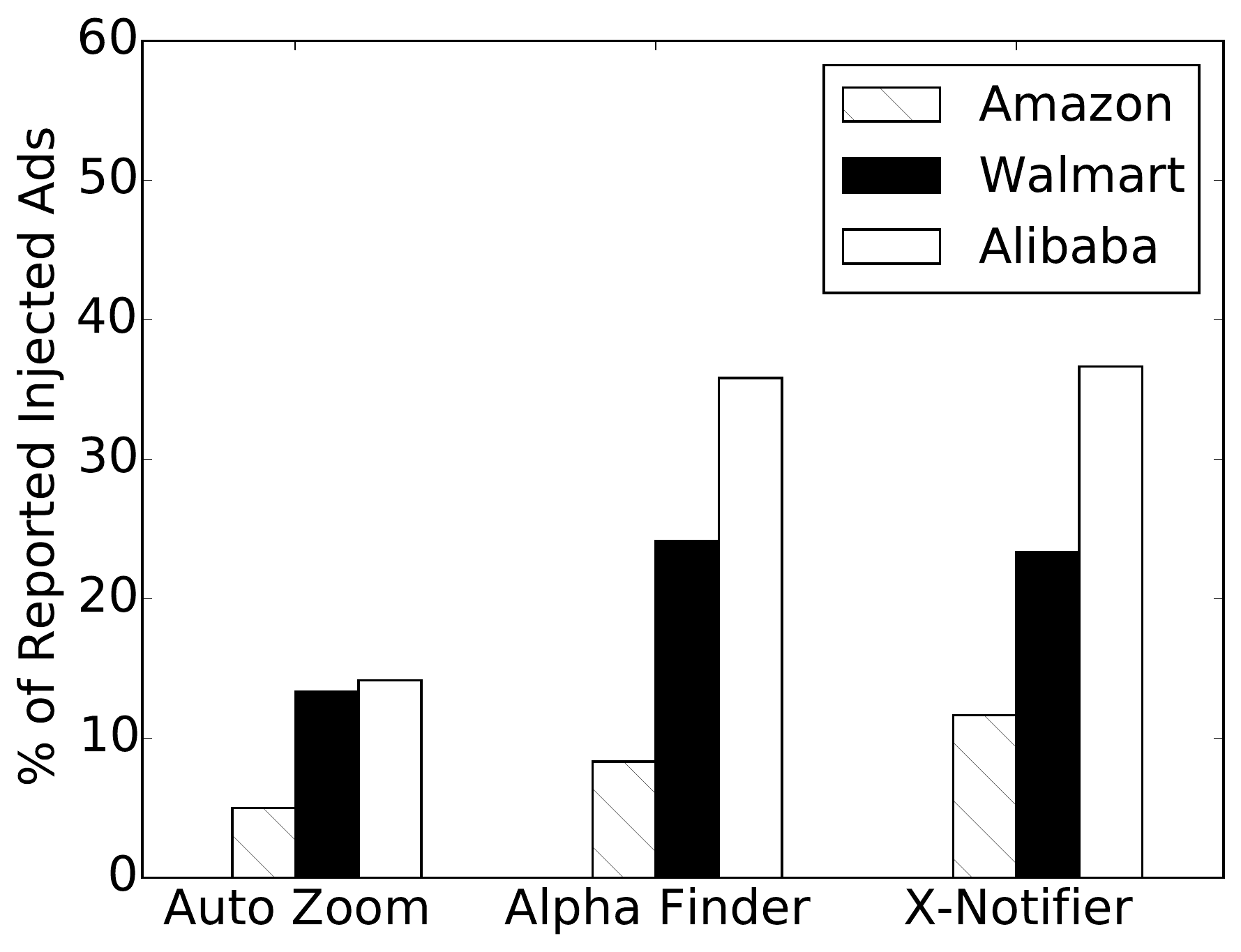}
        \caption{Group 1.}
        \label{fig:eval:user-study:1}
    \end{subfigure}
    \hfill
    \begin{subfigure}[t]{0.48\linewidth}
        \includegraphics[width=\textwidth]{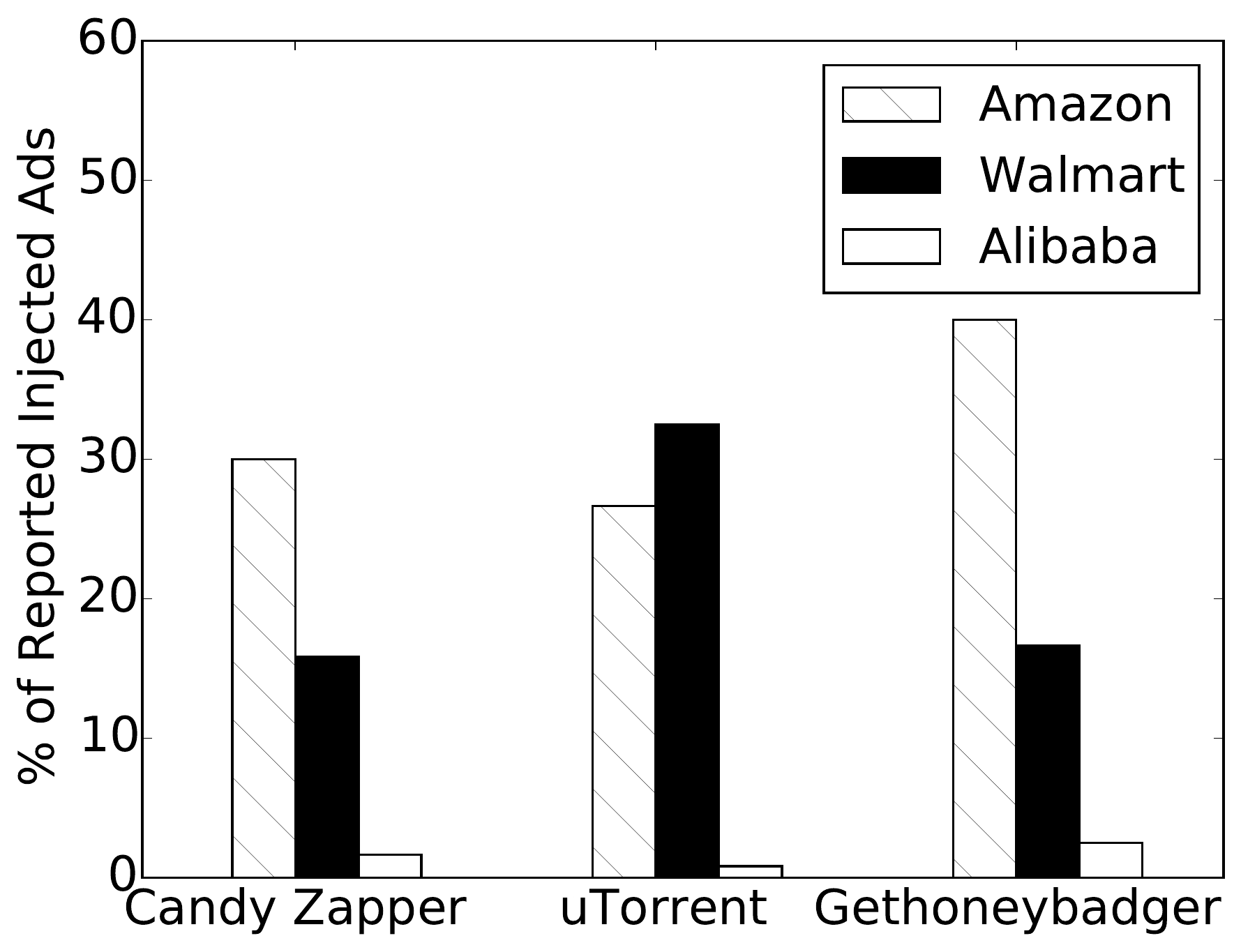}
        \caption{Group 2.}
        \label{fig:eval:user-study:2}
    \end{subfigure}
    \caption{Percentage of injected ads that are reported correctly by all
    the participants.}
    \vspace{-1.5em}
    \label{fig:eval:extensions-adinjection}
    }
\end{figure}

Using each browser, the participants were asked to visit three popular retail
websites: Amazon, Walmart, and Alibaba. Each ad-injecting extension monitors for
visits to these websites, and each injects three different types of
advertisements into these sites. For each website, we asked the participants to
examine the page and tell us if they noticed any content in the page that did
not belong to the website -- in other words, whether any content did not seem to
originate from the publisher. For each group, we aggregated the responses and
presented the percentage of correctly reported ad injection incidents for each
extension in Figure~\ref{fig:eval:extensions-adinjection}.

The results demonstrate that a significant number of Internet users often do not
recognize when ad injection occurs in the wild, even when told to look for
foreign content. For example, 34 participants did not recognize \emph{any}
injected ads out of the three that were added to Amazon website by
\textsf{Auto Zoom} extension. Comparatively more users were able to identify ads
injected by \textsf{Alpha Finder} and \textsf{X-Notifier}. We suspect the reason
for this is because these extensions make use of pop-up advertisements that are
easier to recognize as out-of-place. However, a significant number of users
nevertheless failed to note these pop-up ads, and even after prompting stated
that they thought these ads were part of the publisher's content. More
generally, across all websites and extensions, many participants failed to
identify any injected ads whatsoever.

\begin{figure}[t]
    \centering
    {\sffamily
    \begin{subfigure}[t]{0.32\linewidth}
        \includegraphics[width=\textwidth]{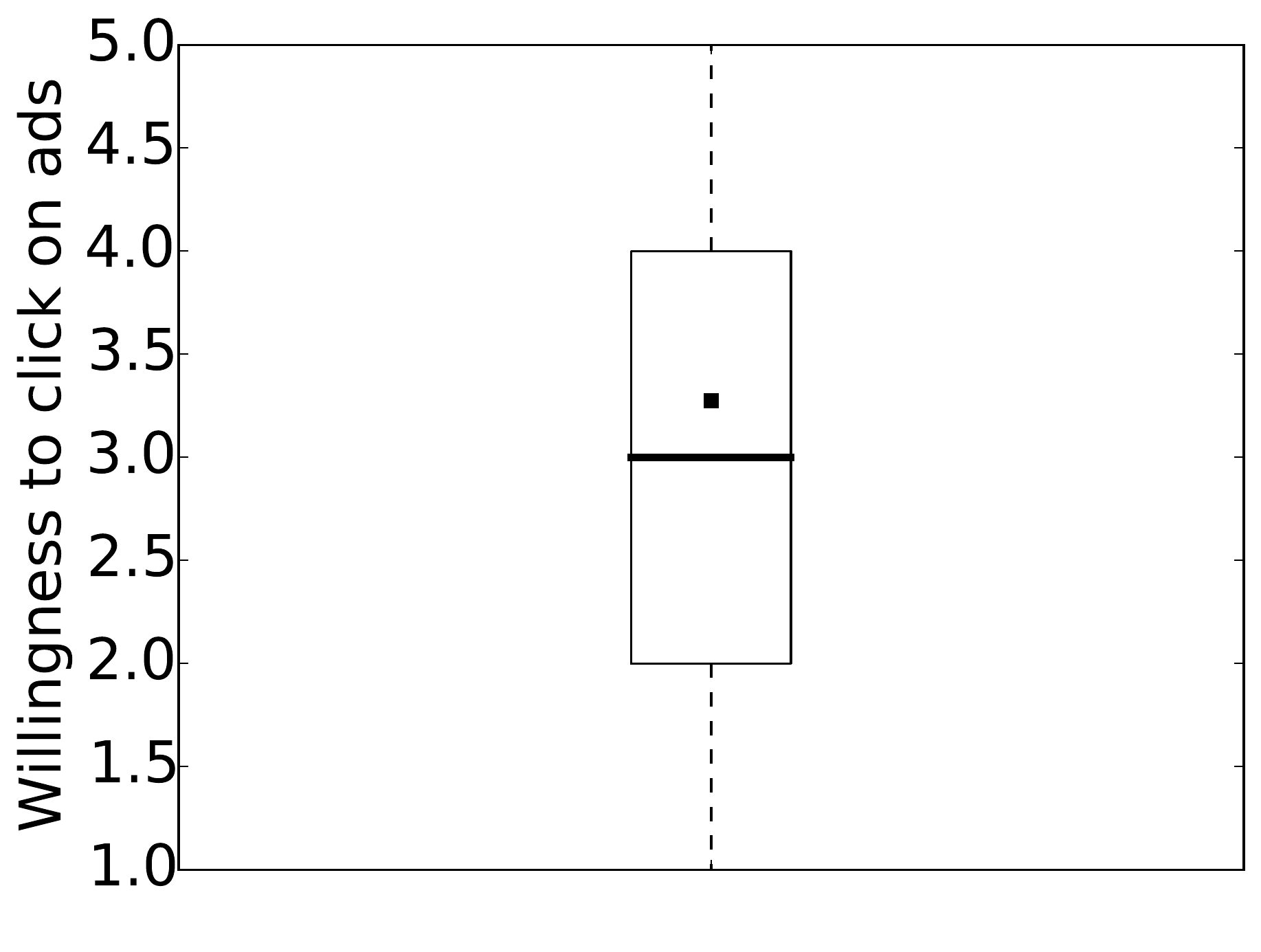}
        \caption{\scriptsize{Susceptibility to ad injection.}}
        \label{fig:eval:user-study:susceptibility}
    \end{subfigure}
    \hfill
    \begin{subfigure}[t]{0.32\linewidth}
        \includegraphics[width=\textwidth]{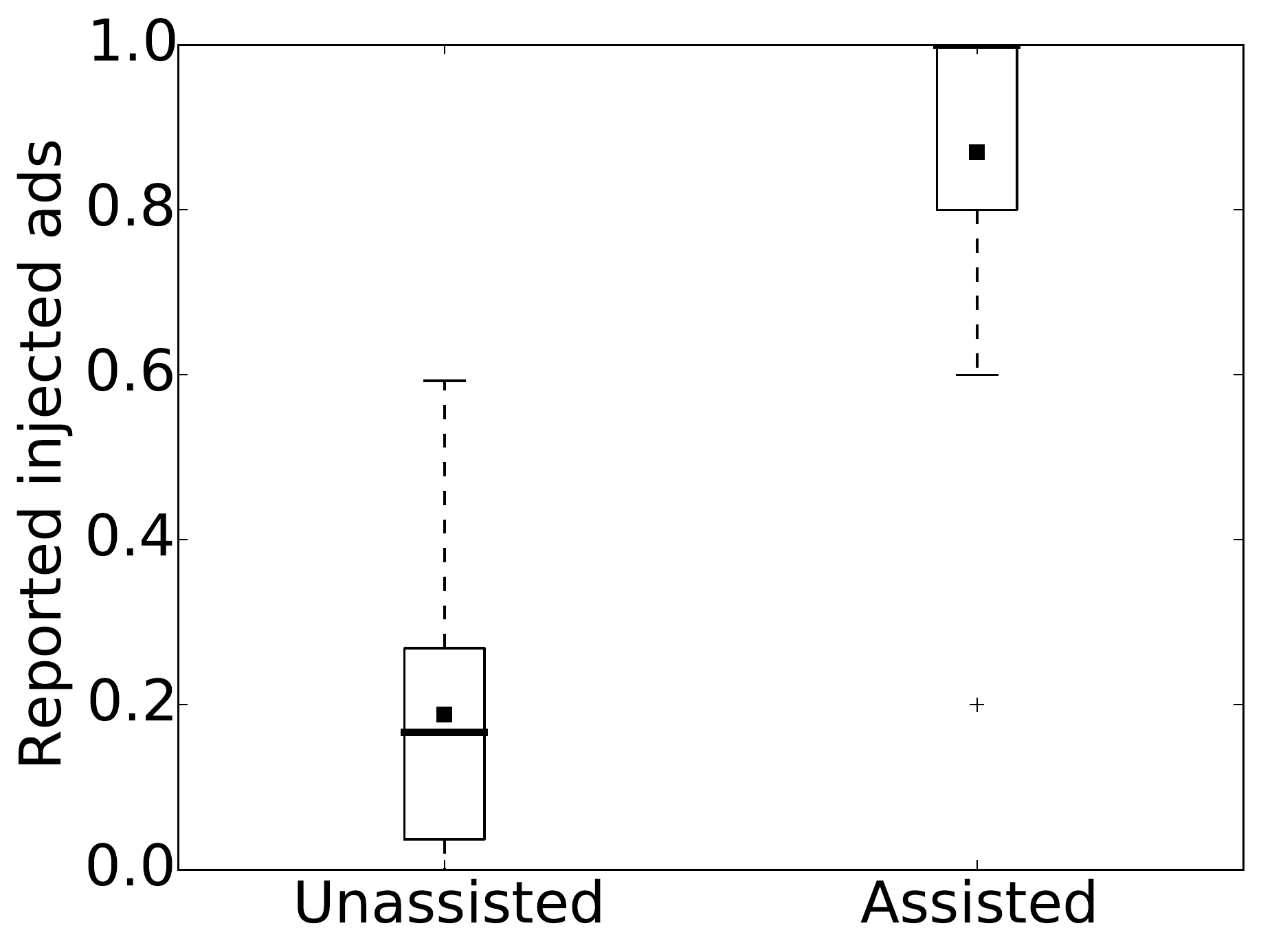}
        \caption{\scriptsize{Ability to identify injected ads.}}
        \label{fig:eval:user-study:identification}
    \end{subfigure}
    \hfill
    \begin{subfigure}[t]{0.32\linewidth}
        \includegraphics[width=\textwidth]{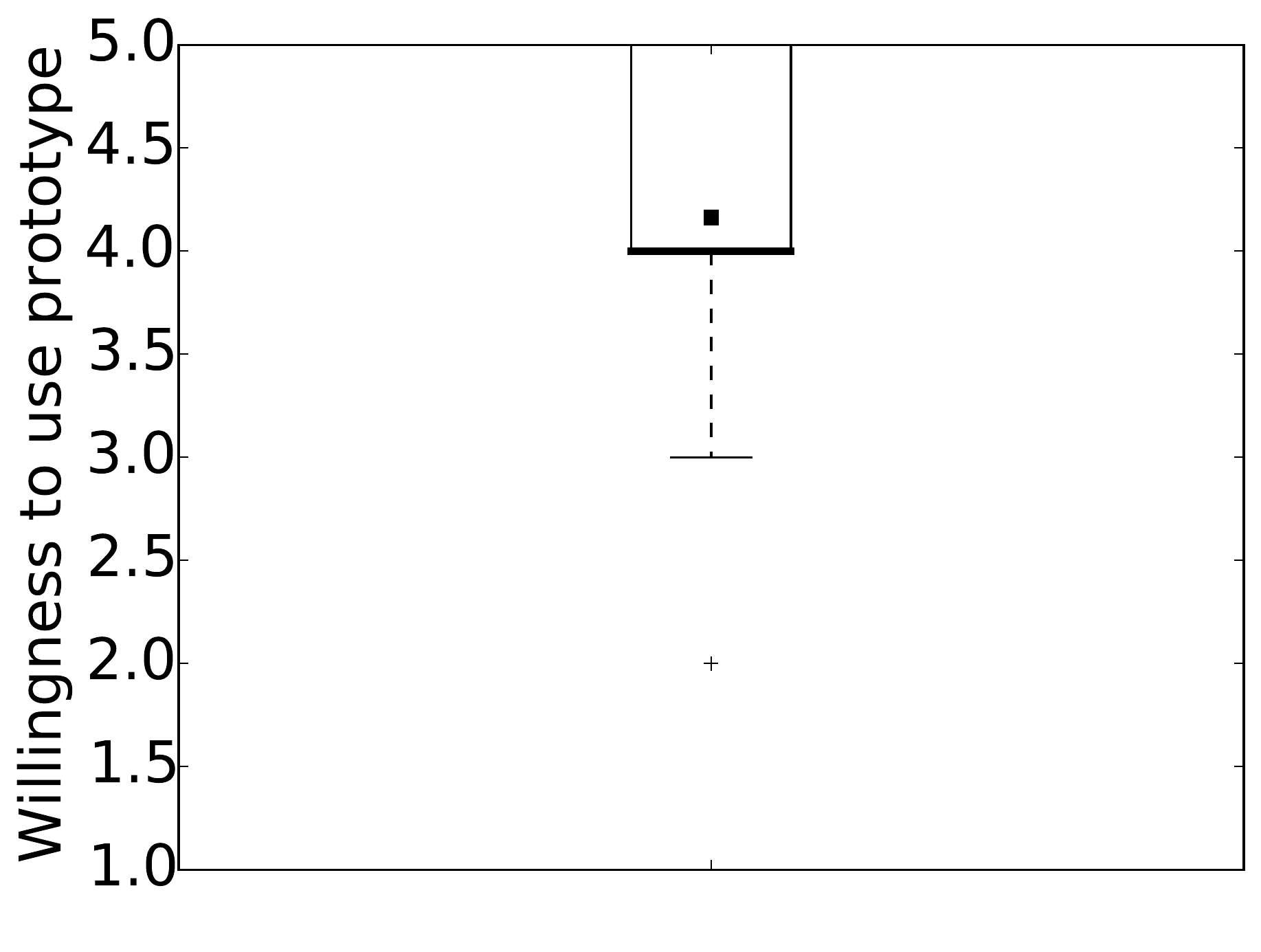}
        \caption{\scriptsize{Usability of content provenance.}}
        \label{fig:eval:user-study:usability}
    \end{subfigure}
    \caption{User study results. For each boxplot, the box represents the
    boundaries of the first and third quartiles. The band within each box
    is the median, while the black square is the mean. The whiskers represent
    1.5 IQR boundaries, and outliers are represented as a \textsf{+} symbol.}
    \vspace{-2.5em}
    \label{fig:eval:user-study}
    }
\end{figure}

We then asked each participant whether they would click on ads in general to
measure the degree of trust that users put into the contents on the publisher's
page. Specifically, we asked participants to rate the likelihood of clicking on
ads on a scale from one to five, where one means that they would never click on
an ad while five means that they would definitely click on an ad. We aggregated
the responses and present the results in Figure~%
\ref{fig:eval:user-study:susceptibility}.

These results show that a significant number of users, roughly half,
\emph{would} click on advertisements that might not originate from the
publisher, but that were instead injected by an extension. This demonstrates the
effectiveness of ad injection as a mechanism for diverting revenue from
publishers to extension authors. It also shows the potential effectiveness of
malicious extensions in using content modifications to expose users to
traditional malware.

\textbf{Effectiveness of Content Provenance Indicators.}
\label{sec:eval:effectiveness}
After the first phase of the experiment, we briefly explained the purpose of
\thesystem and content provenance to the participants. Then, for each
participant in each group, we picked one of the three ad-injecting extensions in
which, the participant did not detect most of the injected ads and installed it
on a Chromium instance equipped with \thesystem. Then, each participant was
asked to visit one of the three retail websites by his choice and identify
third-party content modifications -- i.e., injected ads -- with the help of
provenance indicators. The results are shown in Figure%
~\ref{fig:eval:user-study:identification}, where unassisted identification is
the aggregated number of reported ad injections without any assistance in the
presence of three ad-injecting extensions across three retail websites, and
assisted identification is the number of reported injected ads with the help of
content provenance indicators. Results are normalized to \([0,1]\).

These results clearly imply that users are more likely to recognize the presence
of third-party content modifications using provenance indicators. To confirm
statistical significance, we performed a hypothesis test where the null
hypothesis is that provenance indicators do not assist in identifying
third-party content modifications, while the alternative hypothesis is that
provenance indicators do assist in identifying such content. Using a paired
t-test, we obtain a p-value of \(4.9199 \times 10^{-7}\), sufficient to reject
the null hypothesis at a 1\% significance level. The outliers in assisted
identification are due to the fact that our ad highlighting technique was not
identifiable by a small number of participants. We believe that using different
visual highlighting techniques would make it easier for users to identify the
injected ads.

Finally, we asked each participant how likely they would be to use the content
provenance system in their daily web browsing. We asked participants to rate
this likelihood on a scale from one to five, where one means they would never
use the system and five means that they would always use it. The results are
shown in Figure~\ref{fig:eval:user-study:usability}, and indicate that most
users would be willing to use a content provenance system. The reason behind the
outliers is because a few of the participants stated that they do not need our
system since they would not click on any advertisements. However, we note that
it can be difficult to distinguish between advertisements and other legitimate
content (e.g., products in retail sites) and, consequently, users might be
lured into clicking on ad content injected by extensions.

\textbf{Summary.}
From this user study, we draw several conclusions. First, we confirm that in
many cases users are unable to distinguish injected third-party content from
publisher content. We also show that because users place trust in publishers,
they will often click on injected ads, and thus they tend to be susceptible to
ad injection. Our data shows that content provenance assists in helping users
distinguish between trusted publisher content and injected third-party content
that should not be trusted. Finally, we show that many users would be willing to
use the system based on their experience in this study.

\subsection{Usability}
\label{sec:eval:usability}

We conducted another experiment on a separate population of users to measure the
usability of the \thesystem prototype. The user population was composed of 13
students with different technical background. We presented the participants with
\thesystem integrated into Chromium~43, and asked them to browse the web for
several hours, visiting any websites of their choice. For privacy reasons,
however, we asked users to avoid browsing websites that require a login or that
involve sensitive subject matter (e.g., adult or financial websites). In
addition, for each user, we randomly selected 50 websites from the Alexa Top 500
that satisfy our user privacy constraints and asked the user to visit them. In
particular, each participant was asked to browse at least three levels down from
the home page and visit external links contained in each site. Finally, to gain
some assurance that \thesystem would not break benign extensions, we configured
the browser with the five high-profile extensions list in Table~%
\ref{tab:popular-extensions}.

During the browsing session, the browser was modified to record the number of
URLs visited. We also asked participants to record the number of pages in which
they encountered one of two types of errors. Type~I errors are those where the
browser crashed, system error messages were displayed, pages would not load, or
the website was completely unusable for some other reason. Type~II errors
include non-catastrophic errors that impact usability but did not preclude it --
e.g., the page took an abnormally long time to load, or the appearance of the
page was not as expected. We also asked users to report any broken functionality
for the benign extensions described above as well.

Out of close to 2,000 URLs, two catastrophic errors and 27 non-catastrophic
errors were encountered. However, we note that the majority of URLs rendered and
executed correctly. In addition, none of the participants reported any broken
extensions. We therefore conclude that the proposed approach is compatible with
modern browsers and benign extensions, and further work would very likely allow
the prototype to execute completely free of errors.

\subsection{Performance}
\label{sec:eval:perf}

To measure the performance overhead of \thesystem, we configured both an
unmodified Chromium browser and the prototype to automatically visit the Alexa
Top 1K. The Alexa Top 1K covers many popular websites and is weighted towards
being representative of the sites that people use most often. By using this test
set, we ensured that each browser visited a broad spectrum of websites that
include both static and dynamic content, and especially websites that make heavy
use of third-party components and advertisements. Moreover, we configured both
browser instances with the five benign extensions discussed in Section~%
\ref{sec:background-motivation} that change the DOM to measure performance in
the presence of extensions. A more detailed evaluation would analyze more pages
on these websites to garner a more realistic representation, but that is beyond
the scope of the current work.

We built a crawler based on Selenium WebDriver~\cite{selenium} to automatically
visit the entire list of websites and recorded the total elapsed time from the
beginning of the browsing process until the entire list of websites was visited.
Specifically, our crawler moves to the next website in the list when the current
website is fully loaded, signified by the firing of the \texttt{onload} event.
In order to account for fluctuations in browsing time due to network delays and
the dynamic nature of advertisements, we repeated the experiment 10 times and
measured the average elapsed time. The average elapsed time for browsing the
home pages of the Alexa Top 1K websites measured in this way is 3,457 seconds
for the unmodified browser and 3,821 seconds for \thesystem. Therefore,
\thesystem incurred a 10.5\% overhead on browsing time on average. We also
measured the delay imposed by \thesystem on startup time by launching the
browser 10 times and measuring the average launch time. \thesystem did not cause
any measurable overhead on startup time.

While this overhead is not insignificant, we note that our user study in
Section~\ref{sec:eval:usability} indicates that many users would be willing to
trade off actual perceived performance overhead against the security benefits
provided by the system. Moreover, this prototype is just a proof-of-concept
implementation of our system and there is still room for optimizing the
implementation to decrease the page load time.

\section{Related Work}
\label{sec:related}

\subsection{Malicious Advertising}

Substantial research on malicious advertisements has focused on isolation and
containment~\cite{adsafe,acsac2011adsentry,usenixsec2010adjail}. Other
approaches have focused on detecting drive-by downloads by employing the
properties of HTTP redirections to identify malicious behavior~%
\cite{infocom2014user-browsing-activity,ccs2013spiderweb}. Dynamic analyses have
also been used to detect drive-by downloads and web-hosted
malware~\cite{www2010jsand,sac2010adsandbox,ccs2010blade}. Li et
al.~\cite{ccs2012madtracer} investigated the advertisement delivery process to
detect malvertising by automatically generating detection rules. Web
tripwires~\cite{nsdi2008tripwires} were proposed to detect in-flight page
changes performed by ISPs to inject advertisements.

\subsection{Browser Extension Security}

Browser extension security has recently become a hot topic. The Chromium
extension framework substantially improved the ability of users to limit the
amount of privilege conferred upon potentially vulnerable extensions~%
\cite{tr2008chromium-security-architecture}, and follow-on work has studied the
success of this approach~%
\cite{usenixwebapps2011app-permissions,ndss2012chrome-extensions}. Other work
has broadly studied malicious extensions that attempt to exfiltrate sensitive
user data~\cite{raid2007spyshield,springer2008browserspy}. For instance, Arjun
et al.~showed that many extensions in the Chrome Web Store are over-privileged
for the actual services they purport to provide~\cite{sp2011ibex}.

A recent line of work has focused on the problem of ad injection via browser
extensions. Thomas et al.~\cite{sp2015adinjection} proposed a detection
methodology in which, they used a priori knowledge of a legitimate DOM structure
to report the deviations from that structure as potential ad injections.
However, this approach is not purely client-side and requires cooperation from
content publishers. Expector~\cite{www2015adinjection} inspects a browser
extension and determines if it injects advertisements into websites.
Hulk~\cite{usenixsec2014hulk} is a dynamic analysis system that automatically
detects Chrome extensions that perform certain types of malicious behaviors,
including ad injection. WebEval~\cite{usenixsec2015webeval} is an automatic
system that considers an extension's behaviors, code, and author reputation to
identify malicious extensions distributed through the Chrome Web Store.

In contrast, our work does not attempt to automatically classify extensions that
engage in content modification as malicious or not, but rather focuses on
enabling users to make informed decisions as to whether extensions engage in
desirable behavior or not.

\subsection{Provenance Tracking}

A significant amount of work has examined the use of provenance in various
contexts. For instance, one line of work has studied the collection of
provenance information for generic applications up to entire systems~%
\cite{middleware2012spade,ccs2009sprov,acsac2012hifi}. However, to our
knowledge, no system considers the provenance of fine-grained web content
comprising the DOM. Provenance tracking is also related to information flow
control (IFC), for which a considerable body of work exists at the operating
system level~\cite{sosp2005asbestos,sosp2007flume,nsdi2008dstar}, the language
level~\cite{popl1999jflow,usenixsec2007sif}, as well as the
web~\cite{osdi2012hails,sacmat2010e2eaccctrl}. In contrast to our work, IFC is
focused more on enforcing principled security guarantees for new applications
rather than tracking and indicating data provenance for existing ones.

Numerous systems have examined the use of dynamic taint analysis, a related
concept to provenance. Some prior work~%
\cite{ndss2015info-flows,usenixatc2007spyware-analysis} focuses on tracking
information flow within the browser, Sabre~\cite{acsac2009sabre} detects
whether extensions access sensitive information within the browser, and DSI
enforcement~\cite{ndss2009dsi} defends against XSS attacks by preserving the
integrity of document structure in the browser. While there is certainly an
overlap between dynamic taint analysis and provenance, taint analysis is most
often focused on simple reachability between sources and sinks, while
provenance is concerned with precisely tracking principals that influenced
data.

Finally, there is a line of work that examines provenance on the web. Some
prior work%
~\cite{propr2007provenance-web,ldow2009provenance-web,websci2010provenance-web}
concerns coarse-grained ontologies for describing the origins of data on the
web, and does not consider provenance at a fine-grained scale within the
browser. ESCUDO~\cite{icdcs2010escudo} only considers the principals that can be
controlled by web applications, and it does not handle plug-ins and browser
extensions. LeakTracker~\cite{acns2012leaktracker} performs principal-based
tracking on web pages to study privacy violations related to JavaScript
libraries, but it only tracks injection of scripts into the page, and does not
provide any provenance information for other types of DOM elements.
Excision~\cite{fc2016excision} is the closest work to ours, which tracks
inclusions of different resources in web pages and blocks inclusion of malicious
resources by analyzing inclusion sequences on the page. Although the techniques
are similar, they are used for different purposes. Excision discards the
injection of DOM elements that do not reference remote content (e.g.,
\texttt{div}), and aside from source attributes, that does not track
modifications to DOM elements. However, \thesystem identifies all types of DOM
modification in the page, and instead of blocking content originating from
extensions, it highlights them in the context of the web page.

\section{Conclusion}
\label{sec:conclusion}

In this paper, we introduced fine-grained web content provenance tracking and
demonstrated its use for identifying unwanted third-party content such as
injected advertisements. We evaluated a prototype implementation, a modified
version of Chromium we call \thesystem, through a user study that demonstrated a
statistically significant improvement in the ability of users to identify
unwanted third-party content. Our performance evaluation shows a modest overhead
on a large representative sample of popular websites, while our user experiments
indicate that users are willing to trade off a slight decrease in performance
for more insight into the sources of web content that they browse. We also
performed a comprehensive study on the content modifications performed by
ad-injecting extensions in the wild.

In future work, we plan to explore other uses of provenance on the web. Due to
the highly interconnected structure of the web and the oftentimes obscure nature
of its trust relationships, we believe that surfacing this information in the
form of provenance is a generally useful capability, and can be applied in other
novel ways in order to lead to safer and more informed web browsing. Finally, we
plan to open source our prototype implementation in the hopes that it will be
useful to the wider research community.

\bibliography{paper}{}
\bibliographystyle{plain}

\end{document}